\documentclass{iau}
\usepackage{graphicx}

\title[Are RRab stars fully radial?] %% give here short title %%
{Are RRab stars fully radial?}

\author[J.M. Benk\H{o} \& R. Szab\'o]   %% give here short author list %%
{J\'ozsef M. Benk\H{o}
 \and R\'obert Szab\'o}

\affiliation{Konkoly Observatory, MTA CSFK, \\ Konkoly Thege M. u. 15-17.,
H-1121, Budapest, Hungary \\ email: {\tt benko@konkoly.hu, rszabo@konkoly.hu}
}

\pubyear{2014}
\volume{301}  %% insert here IAU Symposium No.
\pagerange{1--2}
\jname{Precision Asteroseismology}
\editors{J.A. Guzik, W.J. Chaplin, G. Handler \& A. Pigulski, eds.}
\begin{document}

\maketitle

\begin{abstract}
Thanks to the space missions \textit{CoRoT} and \textit{Kepler}
new oscillation frequencies have been
discovered in the Fourier spectra of Blazhko RR
Lyrae stars. The period
doubling (PD) yields half-integer frequencies
between the fundamental mode and its harmonics.
In many cases the first
and/or second radial overtone frequencies also appear temporally. 
Some stars show extra frequencies that were identified as
potential non-radial modes. We show here that
all these frequencies can be explained by pure radial pulsation
as linear combinations of the frequencies of radial
fundamental and overtone modes.
\end{abstract}
%\firstsection % if your document starts with a section,
              % remove some space above using this command.
%\section{The story}

The first RRab star in which non-radial mode
was reported is the \textit{CoRoT} target V1127\,Aql.
\cite[Chadid et al.~(2010)]{Chadid_2010}
explained 468 frequencies detected in this star by four 
independent frequencies, $f_0$, 
$f_{\mathrm m}$, $f'=2f''$, $f_{\mathrm{m1}}$, and their combinations. The
$f_0$ and $f_{\mathrm m}$ mean the main pulsation and modulation
(Blazhko) frequency, respectively, $f'$ (or $f''$) is the frequency of
an independent, possibly non-radial mode, and $f_{\mathrm{m1}}$
is a secondary modulation frequency acting on `additional modes' only
(see \cite[Chadid et al.~2010]{Chadid_2010} for the details).
Later on, frequencies of possible non-radial modes have been
found in the Fourier spectra of the Blazhko stars
CoRoT\,105288363 and V445\,Lyr,
V354\,Lyr and V360\,Lyr observed by \textit{Kepler} 
\cite[(Benk\H{o} et al.~2010 = B10; Guggenberger et al.~2012 = G12)]{Benko_2010,Gugg_2012}.
Many frequencies of these modes were found between the position
of the radial first overtone and the PD
frequencies and yielded period ratios $P/P_0$ around 0.7.

We homogeneously re-analyzed light curves of all the \textit{CoRoT}
and \textit{Kepler} Blazhko RRab stars
in which non-radial mode(s) were reported. 
We used the \textit{CoRoT} 150-days-long data
coadded to get 8-min sampling and \textit{Kepler} 3-years-long 
long-cadence (30-min sampling) data covering Q1-Q12. The
\textit{CoRoT} white fluxes were cleaned and de-trended. In the case
of \textit{Kepler} targets, we used the raw pixel frames
applying our own proper tailor-made apertures for each star and
quarter separately (see \cite[Benk\H{o} et al.~2013]{Benko_2013}).
The data were pre-whitened with the main pulsation
frequencies and their harmonics, the modulation
frequencies and as many modulation side peaks as possible. 
The resulting Fourier spectra and the frequency solutions from 
the literature are compared.

%\section{Frequency solutions with and without non-radial modes}

{\underline {\it V1127\,Aql (CoRoT\,100689962)}}.
Our identification of $f_0$ and $f_{\mathrm m}$ 
are the same as in \cite[Chadid et al.~(2010)]{Chadid_2010}, but if we assume $f'' = f_2-f_0$, then
$f'=2(f_2-f_0)$, where $f_2=4.825397\,$d$^{-1}$ is the frequency of 
the radial second
overtone with the period ratio of 0.582.
This identification eliminates the non-radial mode 
with its period ratio of 0.696.
\cite[Poretti et al.~(2010)]{Poretti_2010} 
have already noticed that
V1127\,Aql shows half-integer frequencies.
If we accept the PD paradigm \cite[(Szab\'o et al.~2010)]{Szabo_2010}, 
the frequency $f_{\mathrm{m1}}$ can also be interpreted as a linear 
combination: 1.5$f_0-f'$.

G12 found the following 
independent frequencies of {\underline {\it CoRoT\,105288363}}: 
$f_0$, $f_{\mathrm m}$,
$f_{\mathrm s}$ (secondary Blazhko frequency),
$f_1$, $f_2$, first 
and second radial overtone modes with the period
ratios 0.745 and 0.590, respectively, and two non-radial modes,
$f_{\mathrm N}= 2.442$~d$^{-1}$
($P_{\mathrm N}/P_0=0.722$) and 
$f_{\mathrm{N2}}= 2.2699$~d$^{-1}$.
The multiple and time-dependent modulation of this star
makes its frequency spectrum complicated. We removed more
side peaks than 
G12, so we
obtained a bit different peak structure. Now, the peak at
$f_{\mathrm{N2}}$ seems to be insignificant while $f_{\mathrm N}$ 
can be identified as $2(f_2-f_0)$.

In the case of
{\underline {\it V445\,Lyr (KIC\,6186029)}} 
G12 reported  
$f_0$, $f_{\mathrm m}$, $f_{\mathrm s}$, 1.5$f_0$ (PD), 
$f_1$, $f_2$ 
% radial overtones with the period ratios 0.73 and 0.585; 
and non-radial mode $f_{\mathrm N}= 2.7719$~d$^{-1}$
($P_{\mathrm N}/P_0=0.703$). 
Many similarities between CoRoT\,105288363 and 
V445\,Lyr have been discussed by G12.
In this study we find an additional one: the previously
suggested non-radial mode $f_{\mathrm N}$ can also be identified as 
$2(f_2-f_0)$.

According to B10, frequency content of 
{\underline {\it V354\,Lyr (KIC\,6183128)}} is the following:
$f_0$, 
$f_{\mathrm m}$, 
$f_2$ ($P_2/P_0=0.586$); 
two independent non-radial modes,
$f'= 2.0810$~d$^{-1}$ and $f'''= 2.6513$~d$^{-1}$, and
$f''= 2.4407$~d$^{-1}$ which was identified 
as a possible radial first overtone mode with the period ratio of 0.729.
The frequency spectrum of the Q1-Q12 data is a bit
different from that for Q1-Q2 data (B10), because the amplitudes of the
additional modes strongly depend on time 
(\cite[B10, Szab\'o~et~al.~2013]{Benko_2010, 
Szabo_2013}) and we removed more
side peaks around harmonics eliminating more aliases.
In consequence, the $f''$ frequency became insignificant. 
We explain $f'= (f_0+f_1)/2$, where 
$f_1=2.3843$~d$^{-1}$ and $f'''=1.5f_0$
(PD). We detected two additional significant peaks at
2.999~d$^{-1}$ and 2.300~d$^{-1}$ which produced an equidistant
triplet with the main PD frequency $f'''$.

Frequency solution for {\underline {\it V360\,Lyr (KIC\,9697825)}} from
B10 is $f_0$, $f_{\mathrm m}$, 
$f_1$ (first overtone mode with the period ratio
$P_1/P_0=0.721$), and $f'=2.6395$~d$^{-1}$,
an independent non-radial mode.
The star shows a consistent picture with the similar stars
if we identify $f'$ and its side peaks as a PD effect, 
and if $f_1$ is identified as $2(f_2-f_0)$, 
where $f_2=3.046$~d$^{-1}$.
%\section{Concluding remarks}

Summarizing:
%\begin{itemize}
%\renewcommand{\labelitemi}{$\bullet$}
%\item
(i) Using linear combination frequencies of radial
modes we obtained alternative solutions for
all those Blazhko RRab stars in which 
non-radial modes were previously suggested. In other words,
our mathematical description explains the spectra solely
by radial modes.

(ii) The amplitudes of the harmonics of combinations 
are many times higher than those of simple combination frequenciesâ
e.g. $A[2(f_2-f_0)] \gg A(f_2-f_0)$. This is unusual but a similar
phenomenon, where the combination frequencies have higher
amplitudes than their components, was reported by 
\cite[Balona et al.~(2013)]{Balona_2013} for a roAp star.
%\item

(iii) We searched for stars which show high-amplitude linear combination 
frequencies in their Fourier
spectra and found at least two additional cases:
CoRoT\,103922434 and V366\,Lyr (KIC\,9578833).
%\end{itemize}

\acknowledgements{
This work was partially supported by the following grants: ESA
PECS No 4000103541 /11/NL/KML, 
Hungarian OTKA Grant K-83790 and KTIA Urkut\_10-1-2011-0019.
RSz acknowledges the J\'anos Bolyai Research Scholarship of the 
Hungarian Academy of Sciences and the IAU travel grant.
}

\end{document}